\documentclass[aps,prd,twocolumn,groupedaddress,superscriptaddress]{revtex4-2}
\usepackage{lipsum}
\usepackage{graphicx}  
\usepackage{dcolumn}   
\usepackage{bm}        
\usepackage{amssymb}   
\usepackage{amsmath}
\usepackage{xcolor}
\usepackage{amsmath}
\usepackage{amsfonts}
\usepackage{hyperref}
\hypersetup{colorlinks=true, linkcolor=blue, citecolor=blue, urlcolor=blue}

\usepackage[normalem]{ulem}

\begin{document}
\preprint{APS/123-QED}
\title{Transport Properties of QGP within a Bayesian Holographic QCD Model}
\author{Bing Chen}
\affiliation{School of Nuclear Science and Technology, University of South China, Hengyang 421001, People's Republic of China}
\author{Liqiang Zhu}
\affiliation{Key Laboratory of Quark and Lepton Physics (MOE) and Institute of Particle Physics, Central China Normal University, Wuhan 430079, China}
\author{Xun Chen}
\email{chenxun@usc.edu.cn}
\affiliation{School of Nuclear Science and Technology, University of South China, Hengyang 421001, People's Republic of China}
\affiliation{INFN -- Istituto Nazionale di Fisica Nucleare -- Sezione di Bari, Via Orabona 4, 70125 Bari, Italy}
\author{Defu Hou}
\email{houdf@mail.ccnu.edu.cn}
\affiliation{Key Laboratory of Quark and Lepton Physics (MOE) and Institute of Particle Physics, Central China Normal University, Wuhan 430079, China}
\author{Xurong Chen}
\email{xchen@impcas.ac.cn}
\affiliation{Institute of Modern Physics, Chinese Academy of Sciences, Lanzhou 730000, People’s Republic of China}

\begin{abstract}
Using a holographic QCD model augmented by Bayesian inference, we calculate key transport coefficients of the quark-gluon plasma (QGP)$\text{--}$including the drag force, jet quenching parameter, heavy quark diffusion coefficient, and shear and bulk viscosities$\text{--}$at finite temperature and chemical potential. Posterior parameter distributions at the 68\% and 95\% confidence levels (CL), as well as the maximum a posteriori (MAP) estimates, are employed to quantify uncertainties. Our findings indicate that the diffusion coefficient within the Bayesian credible regions aligns with lattice QCD results for $T \sim 1.2T_c$ to $2T_c$, and is consistent with ALICE experimental measurements near $T_c$. The jet quenching parameter obtained from the Bayesian analysis agrees with RHIC and LHC data, while viscosity coefficients show compatibility with existing literature. These results demonstrate the efficacy of a Bayesian holographic approach in elucidating the nonperturbative transport properties of QCD matter.
\end{abstract}
\maketitle

\section{Introduction}
\label{sec-int}
The study of high-energy nuclear collisions provides essential insights into the properties of the quark-gluon plasma (QGP), a strongly interacting state of matter believed to have existed shortly after the Big Bang. Among key observables are jet quenching phenomena, characterized by the suppression of high-energy partons traversing the medium, and the transport coefficients governing the medium's response—such as shear viscosity, bulk viscosity, and diffusion constants  \cite{STAR:2005gfr,Yin:2013zea,ATLAS:2010isq,JET:2013cls,CMS:2011iwn}.
The jet quenching parameter $\hat{q}$ quantifies the transverse momentum broadening of energetic partons per unit length, serving as a critical probe of the medium's density and interaction strength \cite{DEramo:2010wup}. Various theoretical models have been developed to calculate this parameter \cite{Wang:1992qdg,Liu:2006ug,Renk:2006sx,Jiang:2022uoe,Jiang:2022vxe,Xing:2024yrb}. 

Shear viscosity $\eta$ measures the ability of a system to restore equilibrium after being subjected to a shear mode perturbation. On a microscopic level, the ratio of shear viscosity to entropy density, $\eta/s$, is closely related to the interaction strength between particles in the system. Typically, a stronger interaction corresponds to a smaller $\eta/s$ ratio. In the weak coupling regime, perturbative QCD calculations show that $\eta \propto 1/(\alpha_s^2 \ln \alpha_s)$ \cite{Arnold:2000dr}, where $\alpha_s$ is the strong coupling constant. When the coupling is strong, Lattice QCD simulations reveal that $\eta/s$ for a purely gluonic plasma remains quite low, generally within the range of 0.1 to 0.2 \cite{Nakamura:2004sy,Ballon-Bayona:2021tzw}. Furthermore, studies based on the AdS/CFT correspondence \cite{Maldacena:1997re,Gubser:1998bc,Witten:1998qj} establish a lower bound for $\eta/s$ as $1/(4\pi)$ \cite{Kovtun:2004de}, a value closely aligned with that employed to fit the elliptic flow $v_2$ data obtained at RHIC \cite{Song:2008hj,Teaney:2000cw,Teaney:2003kp}. Therefore, it is widely recognized that the matter produced at RHIC and LHC behaves as a strongly coupled nearly ``perfect'' fluid.

Bulk viscosity $\zeta$, like shear viscosity, characterizes how quickly a system returns to equilibrium under uniform expansion. In the perturbative regime, $\zeta$ is very small, with its leading dependence on the strong coupling constant $\alpha_s$ expressed as $\zeta \propto \alpha_s^2 / \ln(1/\alpha_s)$ \cite{Arnold:2006fz}. However, studies using various approaches---such as Lattice QCD \cite{Karsch:2007jc,Meyer:2007dy}, the linear sigma model \cite{Paech:2006st}, the Polyakov-loop linear sigma model~\cite{Mao:2009aq}, and the real scalar model~\cite{Li:2008zp}---demonstrate that the ratio $\zeta/s$ rises sharply near the critical temperature $T_c$. This enhancement of bulk viscosity near the phase transition corresponds to the peak observed in the trace anomaly around $T_c$, reflecting a highly non-conformal equation of state~\cite{Boyd:1996bx} in this region. These microscopic insights are further elucidated through complementary studies, utilizing distinct collision systems and probes \cite{Zhang:2025yyd,Wang:2022fwq,Wang:2019vhg,Chen:2024aom,Wang:2013qlv,Xing:2024qcr,Li:2024wqq,Chen:2024eaq,Guo:2024mgh,Du:2024riq}. These approaches provide a multi-scale perspective on QCD matter evolution under extreme conditions.

In strong coupling scenarios, perturbative QCD is inadequate for describing its behavior \cite{Baier:1996kr,Eskola:2004cr}, leading researchers to employ lattice QCD methods to study the static equilibrium properties of this matter. Additionally, there is a non-perturbative approach known as the AdS/CFT correspondence \cite{Maldacena:1997re,Witten:1998qj,Casalderrey-Solana:2011dxg,Probst:2017vsq,Baier:2007ix,Bhattacharyya:2007vjd,Gubser:2008px} that provides a new perspective for investigating the dynamic properties of QGP under strong coupling conditions. This correspondence establishes a connection between $\mathcal{N}=4$ SU($\rm N_c$) super-Yang-Mills theory and type IIB string theory in a combined $\rm AdS_5 \times S^5$ space, offering a powerful tool for analyzing strongly interacting gauge theories when the number of color charges $\rm N_c$ is large and the 't Hooft coupling is also substantial. The original form of this duality linked asymptotically AdS space to a conformal gauge theory at absolute zero temperature. However, since the properties of QGP are closely tied to temperature, researchers have sought to extend this duality to encompass holographic models that describe QGP at non-zero temperatures. This extension has been explored through both top-down \cite{Polchinski:2000uf,Sakai:2004cn,Evans:2011eu,Evans:2010iy,Li:2015uea} and bottom-up \cite{He:2013qq,Yang:2015aia,Gursoy:2007er,Alho:2012mh,Panero:2009tv,Dudal:2015kza,Dudal:2015wfn,Zollner:2024iza,Cao:2024jgt,Wang:2024rim,
Cai:2024eqa,Ahn:2024jkk,Ahn:2024gjf,Jokela:2024xgz,Bea:2024xgv,Zhu:2019ujc,Chen:2024ckb,Chen:2024mmd,Cao:2024jgt} approaches, leading to in-depth investigations in various studies.

One of the intriguing features of QCD is the confinement-deconfinement phase transition,
which is strongly coupled near the phase transition 
\cite{PHENIX:2006gsi,ALICE:2013osk,Matsui:1986dk,Kaczmarek:2005ui,Hashimoto:2014fha}. These phenomena can be explored through the AdS/CFT correspondence and the development of various related holographic dual models \cite{Maldacena:1998im,Rey:1998ik,BitaghsirFadafan:2015zjc,Iatrakis:2015sua,Chen:2017lsf}. Among the holographic models tailored for QCD, those constructed within the framework of Einstein-Maxwell-dilaton (EMD) gravity are particularly noteworthy \cite{DeWolfe:2010he,Chen:2022goa,Rougemont:2023gfz,Jarvinen:2022doa,Knaute:2017opk,Arefeva:2024vom,Dudal:2018ztm,Dudal:2017max,Fu:2024wkn,Li:2025lmp}. In early studies \cite{Gubser:2008yx,Gubser:2008ny}, the authors aimed to construct a five-dimensional gravity theory with black hole solutions to mimic the properties of QCD, introducing an ansatz for the dilaton field potential. These studies indicate that bottom-up holographic models can effectively capture the properties of sound speed and shear viscosity in the quark-gluon plasma. 

Within the framework of the AdS/CFT correspondence, a heavy quark is modeled as a fundamental string attached to a flavor brane. The endpoint of the string corresponds to the quark in the boundary field theory, while the string itself represents the gluonic field surrounding the quark. The resistance encountered by a quark moving through the plasma is reflected in the momentum flux carried from the trailing end of the open string into the deeper regions of the AdS space \cite{Mes:2020vgy,Peng:2024zvf,Gubser:2006qh,Caceres:2006dj,Cheng:2014fza,Gubser:2006bz,Zhu:2019ujc,Domurcukgul:2021qfe,Grefa:2022sav,Giataganas:2013hwa,Zhang:2018mqt,Zhu:2020wds,Andreev:2017bvr,Chen:2023yug,Mykhaylova:2020pfk,Gubser:1996de,Gubser:2007zz,Gubser:2009fc,Gursoy:2009kk,Zhou:2022izh}. 

In this study, we extend a holographic QCD model with Bayesian inference to compute the drag force, jet quenching parameter, heavy quark diffusion coefficient, and viscosities at finite temperature and chemical potential. Employing posterior parameter distributions at 68\%, 95\% confidence levels (CL) and maximum a posteriori (MAP) estimates, we compare our results with lattice QCD, experimental data, and perturbative predictions, thereby providing a comprehensive analysis of the transport properties of the QGP \cite{Rougemont:2015wca,Zhou:2022izh,Zhou:2024oeg,Buchel:2006bv,Herzog:2006gh,Li:2014dsa,Grefa:2023hmf,Apolinario:2022vzg,Sadeghi:2013dga,Wang:2016noh,Du:2023qst,Horowitz:2015dta,Heshmatian:2018wlv,BitaghsirFadafan:2017tci,Li:2014hja,Zhang:2024ebf,Zhang:2023kzf,Hou:2021own,Xing:2021bsc,Li:2020kax,Li:2025tqr,Pokhrel:2024wht,Zhu:2025edv,Li:2025ugv}.

Machine learning is increasingly being applied in high-energy physics \cite{Ma:2023zfj,He:2023zin,Huang:2025uvc,Li:2025csc,Pang:2024kid,Zhou:2023pti,Li:2022ozl,Aarts:2025gyp,Li:2025obt,Soma:2022vbb,Zhang:2025juc,Zhang:2022uqk,Zhou:2018ill,Jiang:2021gsw,Wang:2023exq,Du:2019civ,Du:2020pmp,Du:2021pqa, Du:2025qph,Kou:2024hzd,Kou:2025qsg} and gauge/gravity duality studies \cite{Hashimoto:2018ftp,Hashimoto:2018bnb,Akutagawa:2020yeo, Hashimoto:2021ihd, Hashimoto:2024yev,Song:2020agw,Ahn:2024gjf,Ahn:2024jkk,Ahn:2025tjp,Fu:2024wkn,Cai:2024eqa,Chang:2024ksq,Chen:2024ckb,Chen:2024mmd,Luo:2024iwf,Dai:2025dir,Chen:2025kqb,Guo:2025sbm}. Nowadays, increasing research efforts are being directed toward extracting the specific shear and bulk viscosities of the quark-gluon plasma (QGP) through large-scale model-to-data comparisons combined with Bayesian inference analysis in high-dimensional model parameter space. These efforts also aim to quantify uncertainties arising from variations in other aspects of the theoretical model \cite{Pratt:2015zsa,Bernhard:2016tnd,Bernhard:2019bmu,JETSCAPE:2020mzn,Parkkila:2021yha,Zhu:2025gxo}. Building on a holographic model that captures temperature and chemical potential dependence across QCD phases \cite{Chen:2024ckb,Chen:2024mmd}, previous work calculated the heavy quark potential, Schwinger effect, and transport coefficients \cite{Guo:2024qiq,Zhu:2025xiz,Lin:2024mct,Chen:2024epd,Zhang:2025wxi} and performed Bayesian inference \cite{Zhu:2025gxo}. In this paper, we further calculate QGP transport properties using this holographic framework combined with Bayesian inference.

The paper is organized as follows: Sec. \ref{1} provides a brief review of the holographic QCD model established by the EMD gravity introduced in \cite{Chen:2024ckb,Chen:2024mmd}. Sec. \ref{2} discusses the drag force experienced by a heavy quark in motion within the holographic QCD dynamics model. In Sec. \ref{3}, we calculate the diffusion coefficient of the heavy quark. Sec. \ref{4} is devoted to the examination of the jet quenching parameter.Sec. \ref{5}, we compute the relationship between the bulk viscosity and temperature. Sec. \ref{6} investigates the temperature dependence of the shear viscosity with higher-order derivative corrections.  Sec. \ref{7} presents the overall summary and conclusions of the article.

\section{Holographic Model via Einstein-Maxwell-Dilaton Gravity}
\label{1}
First, we review the five-dimensional EMD system of our model \cite{Chen:2024ckb,Chen:2024mmd,Zhu:2025gxo}. This system comprises a gravitational field $g_{\mu\nu}$, a Maxwell field $A_{\mu}$, and a dilaton field $\phi$. In the Einstein frame, its action is expressed by the following equation:
\begin{equation}\begin{aligned}
&S_E = \frac{1}{16\pi G_5} \\&\int d^5x \sqrt{-g} \left[ R - \frac{f(\phi)}{4} F^2 - \frac{1}{2} \partial_{\mu}\phi\partial^{\mu}\phi - V(\phi) \right].\end{aligned}\label{eq1}
\end{equation}
Here, $R$ is the Ricci scalar, $F_{\mu\nu} = \partial_{\mu}A_{\nu} - \partial_{\nu}A_{\mu}$ is the electromagnetic field tensor, with $f(\phi)$ being the gauge kinetic function coupling to the gauge field $A_{\mu}$, $F$ is the Maxwell field tensor, $V(\phi)$ is the dilaton potential, and $G_5$ is the five-dimensional Newton constant. The explicit forms of the gauge kinetic function $f(\phi)$ and the dilaton potential $V(\phi)$ can be consistently solved through the equations of motion (EOMs).

We propose the following metric ansatz
\begin{equation}
ds^2 = \frac{L^2 e^{2A(z)}}{z^2}\left[-g(z)dt^2 + \frac{dz^2}{g(z)} + d\vec{x}^2\right],\label{eq2}
\end{equation}
where $z$ is the holographic radial coordinate in the fifth dimension and the AdS$_5$ space radius $L$ is conventionally set to one, i.e., $L=1$.

To obtain analytical solutions, we assume the forms of $f(\phi)$ and $A(z)$ along with some parameters. We adopt the metric ansatz
\begin{equation}
A(z) = d \ln(az^2 + 1) + d \ln(bz^4 + 1),\label{eq3}
\end{equation}
and the form of the gauge kinetic function $f(z)$ as
\begin{equation}
f(z) = e^{cz^2 - A(z) + k}.\label{eq4}
\end{equation}
$A(z)$ is set to mimic the correct behavior of entropy and constrain the temperature-dependent model. The function $f(z)$ describes the dependence of model on the chemical potential, which is fixed by the baryon number susceptibility. The constant k corresponds to an overall constant rescaling of the Maxwell term near the conformal boundary. Such a constant prefactor does not change the gauge invariant; it can be absorbed by a field redefinition of the bulk gauge potential.
Then, we can derive
\begin{equation}\begin{aligned}
&g(z) = 1 - \frac{1}{\int_0^{z_h} dx\, x^3 e^{-3A(x)}} \\
&\times\left[\int_0^{z} dx\, x^3 e^{-3A(x)} + \frac{2c\mu^2 e^k}{\left(1 - e^{-cz^2_h}\right)^2} det G\right],\end{aligned} \label{eq5}
\end{equation}
\begin{equation}
\phi'(z) = \sqrt{6 \left(A'^2 - A'' - \frac{2A'}{z}\right)}, \label{eq6}
\end{equation}
\begin{equation}
A_t(z) = \frac{\mu \left(e^{-cz^2} - e^{-cz^2_h}\right)}{1 - e^{-cz^2_h}},\label{eq7}
\end{equation}
and the dilaton potential as
\begin{equation}\begin{aligned}
&V(z) = -\frac{3 z^2 ge^{-2A}}{L^2} \\&\times\left[ A'' + A' \left(3A' - \frac{6}{z} + \frac{3g'}{2g}\right) - \frac{1}{z} \left(- \frac{4}{z} + \frac{3g'}{2g}\right) + \frac{g''}{6g} \right].\end{aligned}\label{eq8}
\end{equation}
The determinant $G$ is given by
\[
\det G = \begin{vmatrix}
\int_0^{z_h} dy\, y^3 e^{-3A(y)} & \int_0^{z_h} dy\, y^3 e^{-3A(y)-cy^2} \\
\int_{z_h}^z dy\, y^3 e^{-3A(y)} & \int_{z_h}^z dy\, y^3 e^{-3A(y)-cy^2}
\end{vmatrix}.
\]
The Hawking temperature and entropy of this black hole solution are given by the following formulas,
\begin{equation}\begin{aligned}
&T = \frac{z_h^3 e^{-3A(z_h)}}{4\pi\int_0^{z_h} dy\, y^3 e^{-3A(y)}} [ 1 + \\&\frac{2c\mu^2 e^k \left( e^{-cz_h^2}\int_0^{z_h} dy\, y^3 e^{-3A(y)} - \int_0^{z_h} dy\, y^3 e^{-3A(y) -cy^2}\right)}{(1 - e^{-cz_h^2})^2}],\end{aligned}\label{eq9}
\end{equation}
\begin{equation}
s = \frac{e^{3A(z_h)}}{4G_5 z_h^3}.\label{eq10}
\end{equation}
For convenient study of our holographic probes of interest, we use the metric in the string frame 
\begin{equation}
ds_s^2 = \frac{L^2 e^{2A_s(z)}}{z^2} \left( -g(z)dt^2 + \frac{dz^2}{g(z)} + dx_1^2 + dx_2^2 + dx_3^2 \right),\label{eq11}
\end{equation}
where $A_s(z) = A(z) + \sqrt{\frac{1}{6}}\phi(z)$.

There are three undetermined parameters $a$, $b$, and $d$ in $A(z)$, and two parameters $c$ and $k$ in $f(z)$, along with the Newton constant $G_5$, making in total a six-dimensional parameter space.

The six parameters $\theta=(a, b, c, d, k, G_5)$ used in our model are extracted from Ref.~\cite{Zhu:2025gxo}. That work employs a Bayesian analysis within the EMD framework, constrained by lattice QCD data for the thermodynamic quantities $(S/T^3, \chi_2^B, C_s^2)$ at zero chemical potential~\cite{HotQCD:2014kol, Bellwied:2015lba}, to determine the posterior distribution of the model parameters. The specific analytical procedure is as follows: first, the parameter inference framework is established based on Bayes' theorem:

\begin{align}
\mathit{P}(\boldsymbol{\theta} | \mathrm{data})\propto\mathit{P}(\mathrm{data} |\boldsymbol{\theta})\mathit{P}(\boldsymbol{\theta})
\end{align}

Within the Bayesian framework, the posterior distribution $\mathit{P}(\theta | \mathrm{data})$ describes the probability distribution of the model parameters $\theta$ after incorporating the observed data. The prior distribution $\mathit{P}(\theta)$ represents the initial belief or existing knowledge about $\theta$ before observing the data. The likelihood function $\mathit{P}(\mathrm{data} | \theta)$ quantifies the plausibility of observing the current data given specific values of $\theta$. Here, "data" specifically refers to lattice QCD data, including $S/T^{3}$, $\chi_{2}^{B}$, and $C_{s}^{2}$, while $\theta$ corresponds to the six parameters in the EMD model.

Regarding the prior distribution $\mathit{P}(\theta)$, the parameter set $\theta=(a,b,c,d,k,G_{5})$ for the EMD model was assigned prior ranges based on existing research~\cite{Chen:2024mmd}, as detailed in Table \uppercase\expandafter{\romannumeral1} of Ref.~\cite{Zhu:2025gxo}. Following these ranges, we generated 300 parameter samples using the Latin Hypercube Sampling(LHS)~\cite{c42ec141-810d-3108-8740-320eb5d0f4b6, MORRIS1995381} method. These parameters were then substituted into the EMD model to compute the corresponding values of $S/T^{3}$, $\chi_{2}^{B}$, and $C_{s}^{2}$. Subsequently, the results were subjected to dimensionality reduction via Principal Component Analysis (PCA)~\cite{10.1162/089976699300016728}. Finally, a Gaussian Process emulator~\cite{williams2006gaussian} was constructed based on the 300 parameter sets and their corresponding PCA-reduced outputs. This emulator will be used to establish the likelihood function in subsequent steps.
\begin{table}[h]
\centering
\begin{tabular}{|c|c|c|}
  \hline
  \multicolumn{3}{|c|}{Prior} \\ \hline
  Parameter & min & max \\ \hline
  $a$ & 0.110 & 0.310 \\ \hline
  $b$ & 0.005 & 0.031 \\ \hline
  $c$ & -0.280 & -0.205 \\ \hline
  $d$ & -0.240 & -0.110 \\ \hline
  $k$ & -0.910 & -0.770 \\ \hline
  $G_{5}$ & 0.375 & 0.430 \\ \hline
\end{tabular}
\caption{The prior parameter ranges for the EMD model~\cite{Zhu:2025gxo}.} 
\label{table1}
\end{table}

Furthermore, for the likelihood function $P(\mathrm{data} \mid \theta)$, we choose to adopt a Gaussian distribution:
\begin{align}
\mathit{P}(\mathrm{\mathrm{data}} |\boldsymbol{\theta})=\underset{i}{\prod}\frac{1}{\sqrt{2\pi}\sigma_{i}}e^{-\frac{[y_{i}(\boldsymbol{\theta})-y_{i}^{\text{lattice}}]^{2}}{2\sigma_{i}^{2}}}
\end{align}

The term $y_{i}^{\text{lattice}}$ represents the lattice QCD results, while $y_{i}(\boldsymbol{\theta})$ is the prediction from the EMD model, which is approximated by a Gaussian emulator to enhance computational efficiency. Here, $\sigma_{i}$ encompasses the uncertainties from both the lattice QCD data and the Gaussian emulator. Finally, by sampling the posterior distribution via the Markov Chain Monte Carlo (MCMC)~\cite{Foreman-Mackey:2012any, Goodman:2010dyf}, we obtained the posterior distributions for the EMD model parameters. The corresponding 95\% Confidence Levels (CL) and the Maximum a Posteriori (MAP) estimates are presented in Table \uppercase\expandafter{\romannumeral2}\cite{Zhu:2025gxo}.

\begin{table}[htbp] 
\centering 
\begin{tabular}{|c|c|c|c|} 
  \hline
  \multicolumn{4}{|c|}{Posterior  95\% CL} \\ \hline
  Parameter & min & max & MAP \\ \hline
  $a$ & 0.229 & 0.282 & 0.252 \\ \hline
  $b$ & 0.019 & 0.027 & 0.023 \\ \hline
  $c$ & -0.261 & -0.231 & -0.245 \\ \hline
  $d$ & -0.143 & -0.127 & -0.135 \\ \hline
  $k$ & -0.871 & -0.808 & -0.843 \\ \hline
  $G_{5}$ & 0.388 & 0.406 & 0.397 \\ \hline
\end{tabular}
\caption{In the 2+1 flavor QCD system, the distributions of the EMD model parameters $(a, b, c, d, k, G_{5})$ after Bayesian inference, including the maximum a posteriori values and the 95\% CL ranges\cite{Zhu:2025gxo}.} 
\label{table:parameter}
\end{table}

\section{Heavy Quark Drag Force}
\label{2}
\begin{figure*}
    \centering
    \includegraphics[width=0.95\textwidth]{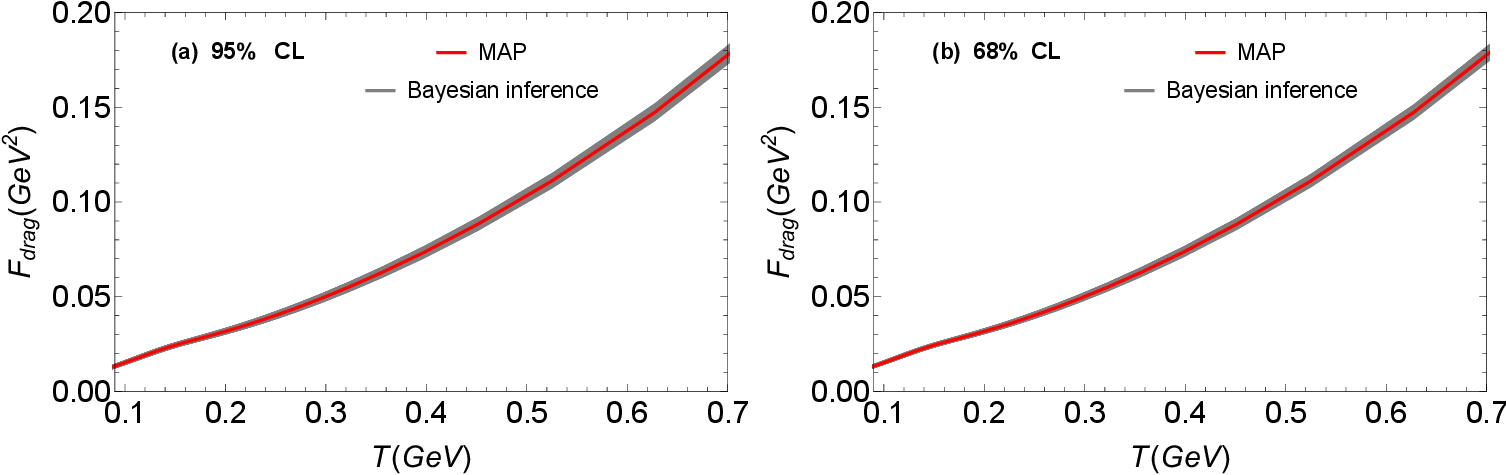}
    \caption{\label{F1} In a 2+1 flavor system at zero chemical potential with the quark velocity set to \( v = 0.3 \), drag force as a function of temperature is examined. (a) The gray area represents the variation of drag force with temperature at the 95\% CL; (b) the gray area represents the variation of drag force with temperature at the 68\% CL. The red curve shows the variation of drag force with temperature at the MAP values.
}
\end{figure*}

In the holographic trailing string approach, when a heavy quark moves through a strongly coupled medium at a constant velocity \(v\) in some direction (for example, in the \(x\) direction), it is represented by an endpoint of an open string attached to the boundary, while the remainder of the string trails behind it, with the other end connected to a new two-dimensional black hole horizon that develops over the string world sheet within the bulk. As the quark moves at the boundary, it loses energy and momentum through the drag force \(F_{\text{drag}} = \frac{d p_x}{dt}\), which can be computed through the energy flow \(\frac{dE}{dx}\) from the string endpoint at the boundary to the other string endpoint located at the world sheet horizon within the bulk \cite{Rougemont:2015wca}.
For the EMD model, it has been shown in \cite{Chen:2024epd} that the heavy quark drag force is given by the following formula
\begin{equation}\begin{aligned}
F_{\text{drag}} =\frac{dp}{dt}=\frac{dE}{dx}= -\pi_x^1= -\frac{1}{2\pi\alpha'} \frac{L^2 e^{2A_s(z_s)}v}{ z_s^2},
\end{aligned}\label{eq12}
\end{equation}
where $\sqrt{\lambda} = \frac{g_{YM}^2 N_c}{4\pi} = \frac{L^2}{\alpha'}$ is the 't Hooft coupling. The only new parameter, $\alpha'$, is set to 2. Its value is determined by fitting to the NLO perturbative results in the high-temperature region and by minimizing the corresponding deviation. \( z_s \) is the radial location of the string world sheet horizon \cite{Rougemont:2015wca}, which is obtained as the numerical solution of the following equation \cite{Chen:2024epd}
\begin{equation}
g(z_s) - v^2 = 0.\label{eq13}
\end{equation}
The drag force in the AdS/Schwarzchild background can be obtained as follows \cite{Gubser:2006qh},
\begin{equation}
F_{\text{drag}}^{\text{SYM}} = -\frac{\pi T^2\sqrt{\lambda}}{2} \frac{v}{\sqrt{1 - v^2}}.\label{eq22}
\end{equation}
In Fig. \ref{F1}, we study the variation of drag force with temperature in the 2+1 flavor system at zero chemical potential and the quark velocity set to \( v = 0.3 \). The gray region in Fig. \ref{F1} (a) represents the range of drag force under 95\% CL, with the red curve showing the change in drag force with temperature at the MAP values. In Fig. \ref{F1} (b), the gray region represents the range of drag force under 68\% CL, with the red curve again depicting the change in drag force with temperature at the MAP values. The figure clearly shows that the resistance increases significantly as the temperature rises.

\begin{figure*}
    \centering
    \includegraphics[width=0.95\textwidth]{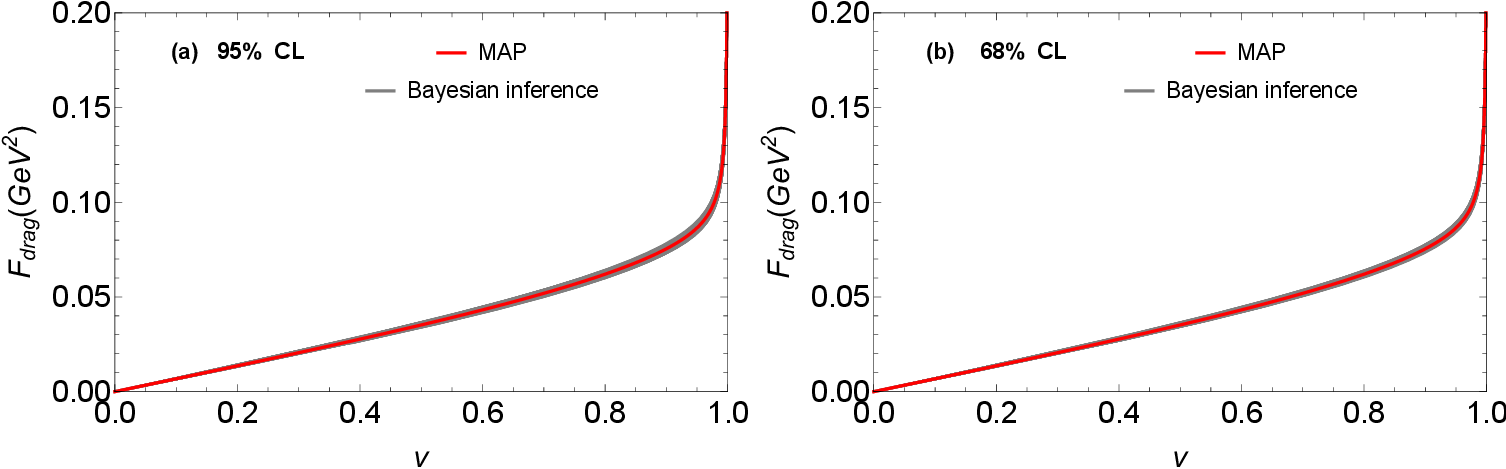}
    \caption{\label{F2} In a 2+1 flavor system at zero chemical potential and at the critical temperature \( T_c = 0.128 \) GeV, drag force as a function of velocity is investigated. (a) The gray area represents the variation of drag force with velocity at the 95\% CL; (b) the gray area represents the variation of drag force with velocity at the 68\% CL. The red curve shows the variation of drag force with velocity at the MAP values.
}
\end{figure*}

In Fig. \ref{F2}, the variation of drag force with velocity in the 2+1 flavor system at the critical temperature \(T_c = 0.128 \, \text{GeV}\) and zero chemical potential is presented. The gray region in Fig. \ref{F2} (a) represents the range of drag force under 95\% CL, while the red curve shows the change in drag force with velocity at the MAP values. In Fig. \ref{F2} (b), the gray region represents the range of drag force under 68\% CL, with the red curve again depicting the change in drag force with velocity at the MAP values. The figure clearly indicates that the drag force increases with velocity, suggesting that in the 2+1 flavor system, the resistance faced by a moving quark increases as its velocity rises. The higher the velocity, the greater the medium resistance that the quark needs to overcome.

\begin{figure*}
    \centering
    \includegraphics[width=0.93\textwidth]{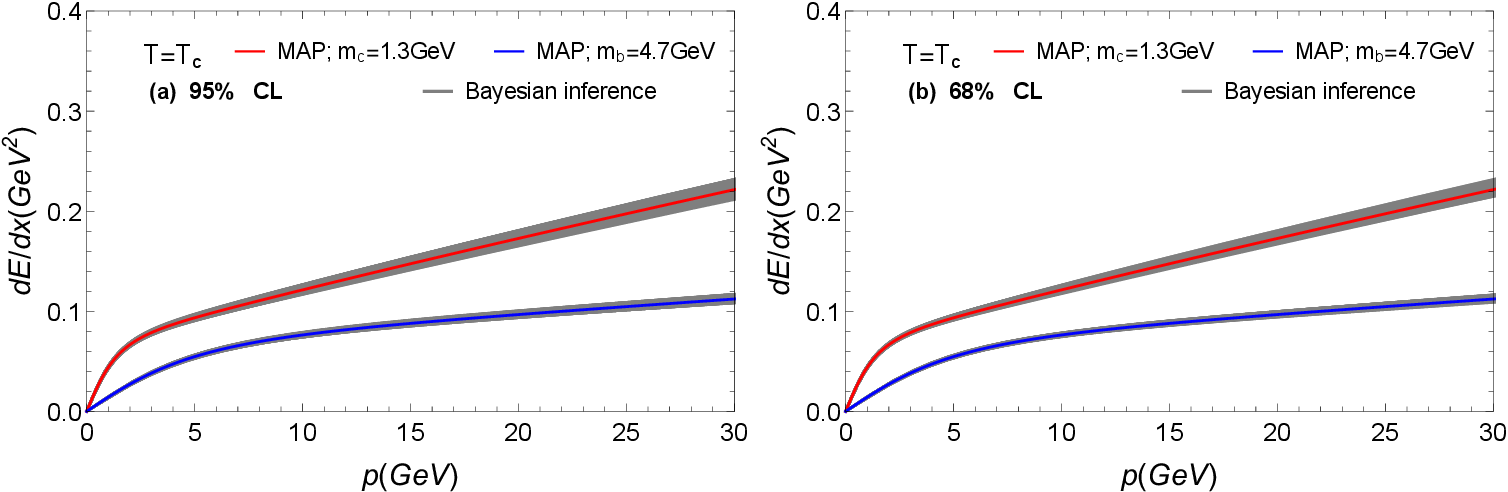}
    \caption{\label{E1} Energy loss of the bottom quark ($m_b = 4.7 \enspace \rm GeV$) and charm quark ($m_c = 1.3 \enspace \rm GeV$) as a function of momentum $p$ (in GeV) at temperature \(T=T_c\). (a) The gray area represents the relationship between energy loss and momentum at the 95\% CL. (b) The gray area represents the relationship between energy loss and momentum at the 68\% CL. The red and blue curve shows the results at the MAP values.
}
\end{figure*}

\begin{figure*}
    \centering
    \includegraphics[width=0.93\textwidth]{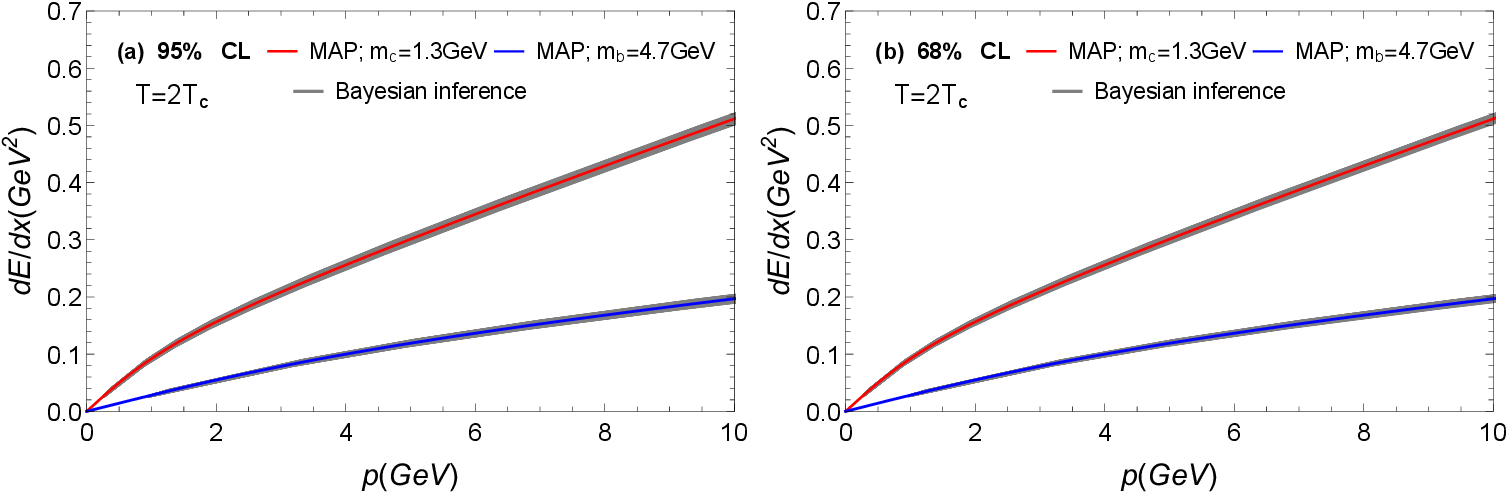}
    \caption{\label{E2} Energy loss of the bottom quark ($m_b = 4.7 \enspace \rm GeV$) and charm quark ($m_c = 1.3 \enspace \rm GeV$) as a function of momentum $p$ (in GeV) at temperature \(T=2T_c\). (a) The gray area represents the relationship between energy loss and momentum at the 95\% CL. (b) The gray area represents the relationship between energy loss and momentum at the 68\% CL. The red and blue curve shows the results at the MAP values.
}
\end{figure*}

According to Eq. (\ref{eq12}), it can be derived that energy loss is equal to the drag force, which allows us to plot the relationship between energy loss and momentum in Bayesian inference. Fig. \ref{E1} illustrates the relationship between the bottom quark ($m_b = 4.7\enspace \rm GeV$) and charm quark ($m_c = 1.3\enspace \rm GeV$) in a 2+1 flavor system at zero chemical potential and temperature \(T=T_c\) \cite{Guo:2024mgh,Du:2024riq}. Fig. \ref{E2} shows the relationship between the bottom quark and charm quark in a 2+1 flavor system at zero chemical potential and temperature \(T=2T_c\), where the light-colored area represents the range of energy loss under 68\% CL and 95\% CL, and the curve line indicates the energy loss value at the MAP values. Fig. \ref{E1} (a) and  Fig. \ref{E1} (b) represent the 95\% CL and 68\% CL, respectively, and the same applies to Fig. \ref{E2}. From Fig. \ref{E1} and Fig. \ref{E2}, it is evident that energy loss increases with momentum. The mass of the quarks also affects energy loss; lighter quarks result in greater energy loss. Additionally, a comparison between Fig. \ref{E1} and Fig. \ref{E2} shows that temperature has a more significant impact on energy loss than quark mass, with higher temperatures leading to increased energy loss.

\section{Heavy Quark Diffusion Coefficient}
\label{3}
In the AdS/Schwarzchild background, the drag force Eq. (\ref{eq22}) can be rewritten as  \cite{Gubser:2006qh}
\begin{equation}
F_{\text{drag}}^{\text{SYM}} = -\frac{\pi T^2 \sqrt{\lambda}}{2m} \frac{v m}{\sqrt{1 - v^2}} = -\eta_D p, \label{eq23}
\end{equation}
where \( m \) denotes the mass of the heavy quark, \( \eta_D \) is the drag coefficient, and \( p = \frac{vm}{\sqrt{1 - v^2}} \) is the momentum.
The diffusion time \( t_{\text{SYM}} \) is given by  \cite{Gubser:2006qh}
\begin{equation}
t_{\text{SYM}} = \frac{1}{\eta_D} = \frac{2m}{\pi T^2 \sqrt{\lambda}}, \label{eq24}
\end{equation}
and the diffusion coefficient \( D_{\text{SYM}} \) can be expressed as  \cite{Gubser:2006qh}
\begin{equation}
D_{\text{SYM}} = \frac{T}{m} t_{\text{SYM}} = \frac{2}{\pi T \sqrt{\lambda}}. \label{eq25}
\end{equation}
Eq. (\ref{eq12}) can be rewritten as
\begin{equation}
F_{\text{drag}} = -\frac{e^{2A_s(z_s)}\sqrt{1 - v^2}}{\pi^2 T^2 z_s^2} \frac{\pi T^2 \sqrt{\lambda}}{2m} \frac{vm}{\sqrt{1 - v^2}}, \label{eq26}
\end{equation}
The diffusion time \( t \) is
\begin{equation}
t = \frac{2m}{\pi T^2 \sqrt{\lambda}} \frac{\pi^2 T^2 z_s^2}{e^{2A_s(z_s)} \sqrt{1 - v^2}}. \label{eq27}
\end{equation}
The diffusion coefficient \( D \) can be represented as
\begin{equation}
D = \frac{T}{m} t = \frac{2}{\pi T \sqrt{\lambda}} \frac{\pi^2 T^2 z_s^2}{e^{2A_s(z_s)} \sqrt{1 - v^2}}. \label{eq28}
\end{equation}
From Eq. (\ref{eq25}) and Eq. (\ref{eq28}), we can deduce
\begin{equation}
\frac{D}{D_{\text{SYM}}} = \frac{\pi^2 T^2 z_s^2}{e^{2A_s(z_s)} \sqrt{1 - v^2}}. \label{eq29}
\end{equation}

\begin{figure*}
    \centering
    \includegraphics[width=0.95\textwidth]{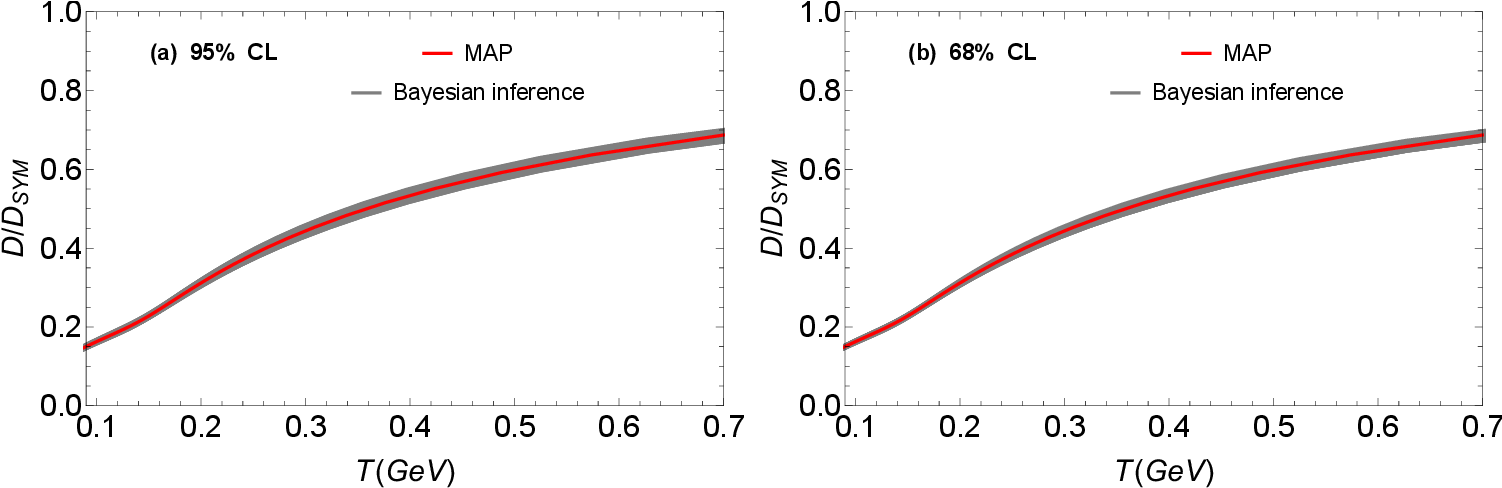}
    \caption{\label{D}  In the 2+1 flavor system at zero chemical potential, with the quark velocity set to \(v = 0.3\), the variation of the scaled diffusion coefficient \(D/D_{SYM}\) with temperature can be described as follows. (a) The gray area represents the variation of \(D/D_{SYM}\) with temperature at the 95\% CL
    ; (b) the gray area represents the variation of \(D/D_{SYM}\) with temperature at the 68\% CL. The red curve shows the variation of \(D/D_{SYM}\) with temperature at the MAP values.
}
\end{figure*}

From Fig. \ref{D}, it can be observed that in the 2+1 system with a quark velocity of \( v = 0.3 \) and a chemical potential of zero, as the temperature increases, the ratio of the diffusion coefficient to its conformal value in the \(\mathcal{N}=4\) SYM theory is also increasing. This growth indicates that the diffusive properties of the system become increasingly aligned with the behaviors predicted by conformal field theory at higher temperatures. The figure also shows that under zero chemical potential, this ratio ultimately trends towards 1 as the temperature rises. The gray area in the graph represents \(D/D_{SYM}\) at the 68\% CL and 95\% CL , while the red curve indicates the \(D/D_{SYM}\) at the MAP values.

\begin{figure*}
    \centering
    \includegraphics[width=0.93\textwidth]{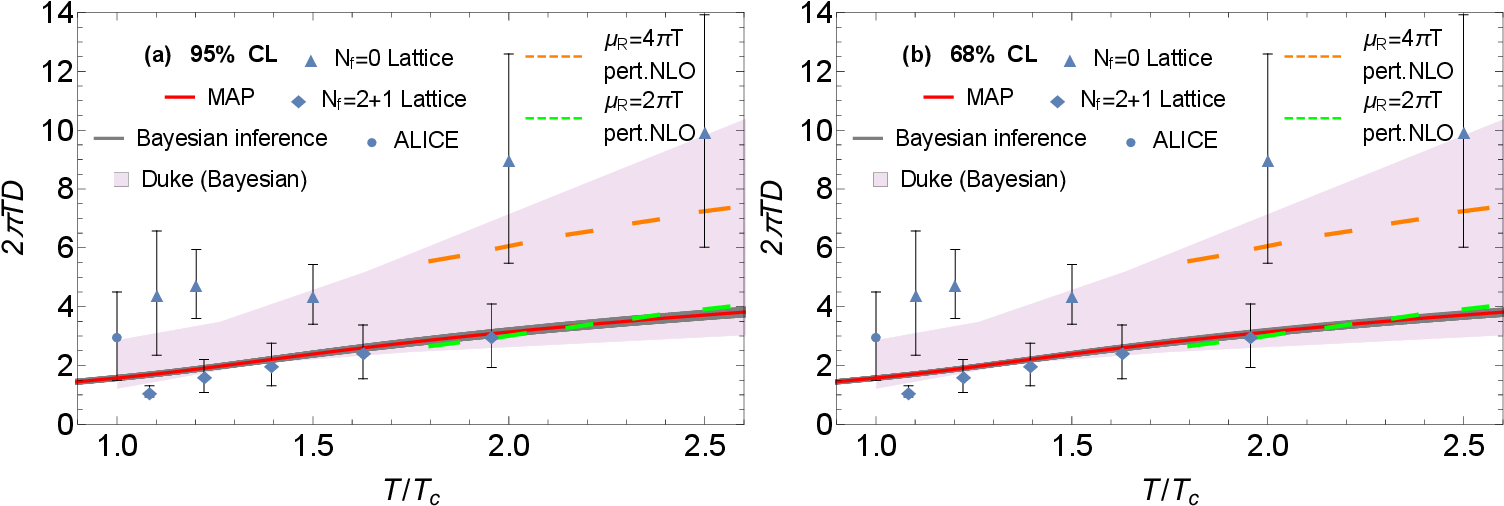}
    \caption{\label{b} The scaled diffusion coefficient $2\pi TD$ as a function of $T/T_c$ , with the quark velocity set to \(v = 0.3\). The lattice data in the figure are results from Ref. \cite{Altenkort:2023oms}. The results of ALICE are from Ref. \cite{ALICE:2021rxa}.The orange dashed line represents the next-order perturbative calculation with the renormalization scale set to \( \mu_R = 4\pi T \), while the green dashed line represents the next-order perturbative calculation with the renormalization scale set to \( \mu_R = 2\pi T \) \cite{Caron-Huot:2007rwy}. The pink region represents Bayesian inference results from the Duke hydro/transport model \cite{Xu:2017obm}. The red curves and gray area represents the results of our model. (a) The gray area represents the variation of \(2\pi TD\) with  $T/T_c$ at the 95\% CL; (b) the gray area represents the variation of \(2\pi TD\) with  $T/T_c$ at the 68\% CL. The red curve shows the variation of \(2\pi TD\) with temperature at the MAP values.
}
\end{figure*}

We compared the spatial heavy quark diffusion coefficient, normalized by \(2\pi T\), with estimates from lattice QCD, ALICE experiments, and Next-to-Leading Order (NLO) perturbative predictions, as depicted in Fig. \ref{b}, it can be seen that our model's 68\% CL, 95\% CL and MAP values results almost fall within the error bars of the lattice data for \(N_f = 2+1\) \cite{Altenkort:2023oms}. 
The gray uncertainty band represents the computed results using the posterior parameters at the 68\%CL and 95\%CL, while the red curve corresponds to the result obtained from the MAP parameter set. The uncertainty in the gray band originates from the range of the posterior parameters.
Our results are also in very good agreement with the results from the ALICE experiment \cite{ALICE:2021rxa}. At high temperatures, our values of \(2\pi TD\) are basically consistent with the NLO perturbative predictions \cite{Caron-Huot:2007rwy}. Additionally, our Bayesian inference completely coincide with the region of the Duke hydro/transport model \cite{Xu:2017obm}.

\section{Jet Quenching Parameter}
\label{4}
We now turn to the study of the jet quenching parameter. Jet quenching measures the energy loss rate of energetic partons as they traverse the created hot dense medium. In Ref. \cite{Chen:2024epd}, the authors conducted a detailed study on the temperature-dependent behavior of \( \hat{q} \) within this model.

According to Ref. \cite{Liu:2006ug}, the relationship between the jet quenching parameter and the adjoint light-like Wilson loop is given by the  equation
\begin{equation}
\langle W^A[C] \rangle \approx \exp\left(-\frac{1}{4\sqrt{2}} \hat{q} L^{-} L'^2\right), \label{eq30}
\end{equation}
where $W^A[C]$ is the Wilson loop in the adjoint representation, and $C$ is a rectangular contour of size $L' \times L^{-}$. The quark and antiquark are separated by a small $L'$ and travel along the $L^-$.

The expectation value of the Wilson loop is dual to the classical value of the string Nambu-Goto action under the appropriate string configuration. The \( \hat{q} \) can be obtained as \cite{Chen:2024epd}
\begin{equation}
\hat{q} = \frac{1}{\pi \alpha' \int_0^{z_h} \frac{dz \, z^2 L^{-2} e^{-2A_s(z)}}{ \sqrt{g(z) (1 - g(z))}}}.\label{eq31}
\end{equation}
For $\mathcal{N}=4$ supersymmetric Yang-Mills theory, in the large \( N_c \) and large \( \lambda \) limit, Eq. (\ref{eq31}) leads to the following analytic expression \cite{Liu:2006ug}
\begin{equation}
\hat{q}_{SYM} = \frac{\pi^{3/2} \Gamma\left(\frac{3}{4}\right)}{\Gamma\left(\frac{5}{4}\right)} \sqrt{\lambda}T^3, \label{eq46}
\end{equation}
where \( \Gamma \) denotes the Gamma function.  The finite 't Hooft coupling correction on this jet quenching parameter was obtained in Ref. \cite{Zhang:2012jd}, and the jet quenching parameter was found to be reduced due to world sheet fluctuations by a factor
$( 1-1.97\lambda^{-1/2})$.

\begin{figure*}
    \centering
    \includegraphics[width=0.95\textwidth]{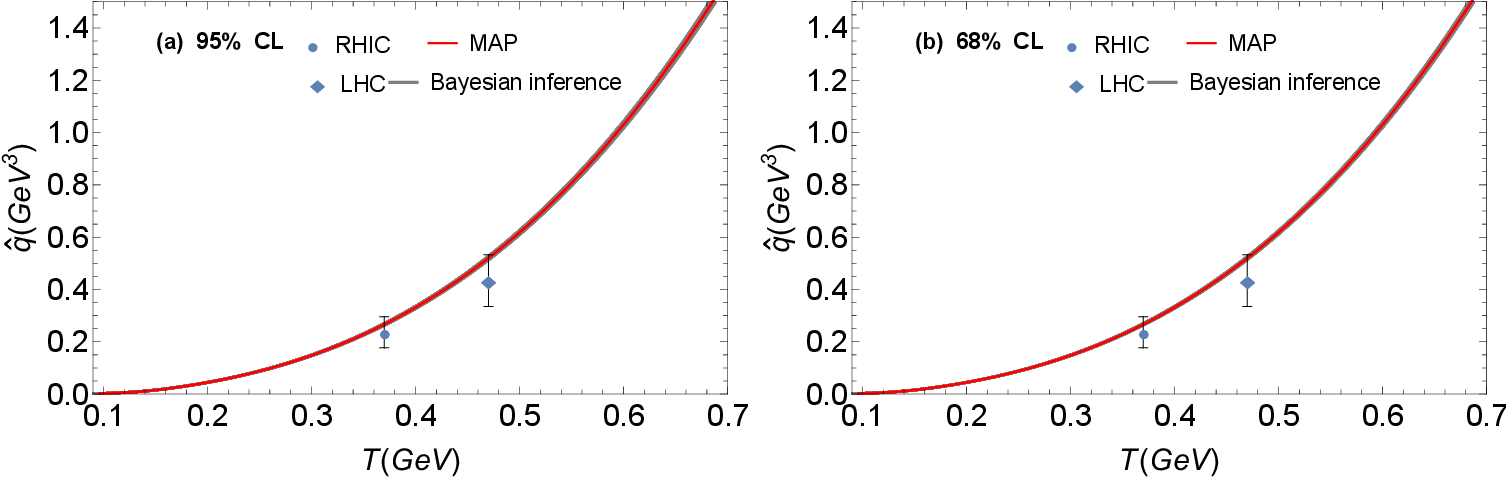}
    \caption{\label{q}   The relationship between the jet quenching parameter and temperature in the holographic model at zero chemical potential. Error bars represent experimental values from RHIC and LHC \cite{JET:2013cls}. (a) The gray area represents the variation of jet quenching parameter with temperature at the 95\% CL; (b) the gray area represents the variation of jet quenching parameter with temperature at the 68\% CL. The red curve shows the variation of jet quenching parameter with temperature at the MAP values.
}
\end{figure*}

To obtain the jet quenching parameters in the holographic QCD model, we performed numerical calculations at various temperatures with zero chemical potential. The resulting curves are illustrated in Fig. \ref{q}. It can be observed that the increase in temperature leads to an enhancement of the jet quenching parameter. This indicates that in the model considered, the medium is denser or hotter, resulting in increased energy loss. This is consistent with the physical intuition that jets passing through a medium at higher temperatures will encounter more scattering centers and, hence, experience greater energy loss. Our model's Bayesian inference are consistent with the experimental results from RHIC and LHC \cite{JET:2013cls}.

\begin{figure*}
    \centering
    \includegraphics[width=0.95\textwidth]{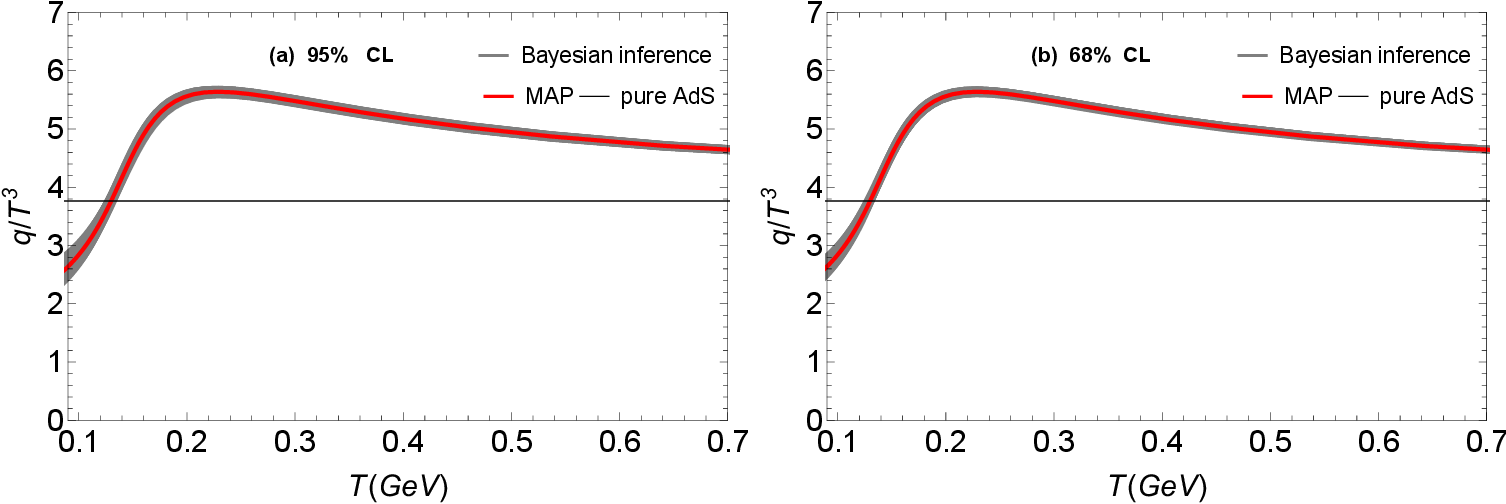}
    \caption{\label{qT}  The relationship between \(\hat{q}/T^3\) and temperature \(T\) at zero chemical potential. (a) The gray area represents the variation of $\hat{q}/T^3$ with temperature at the 95\% CL; (b) the gray area represents the variation of $\hat{q}/T^3$ with temperature at the 68\% CL. The red curve shows the variation of $\hat{q}/T^3$ with temperature at the MAP values.
}
\end{figure*}

\begin{figure*}
    \centering
    \includegraphics[width=0.94\textwidth]{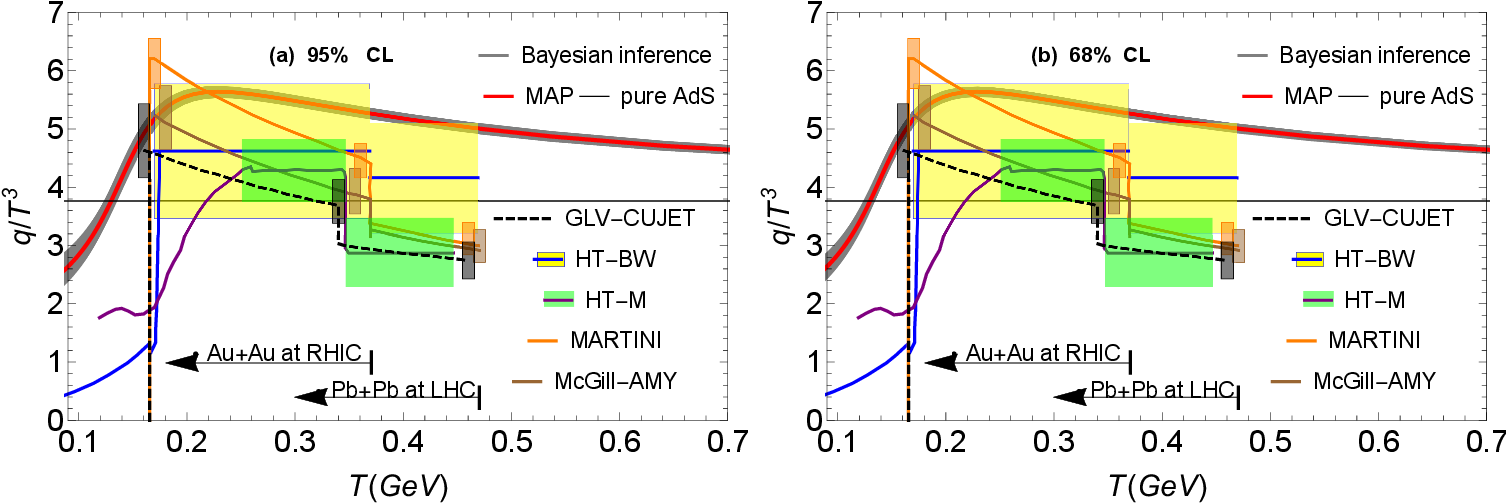}
    \caption{\label{qTT}  The relationship for $\hat{q}/T^3$ at zero chemical potential with temperature $T$. The plot (HT-BW and HT-M) of the scaled jet transport parameter \(\hat{q}/ T^3\) as a function of the initial temperature \(T\) for an initial quark jet with energy \(E = 10\) GeV in the most central A+A collisions, with an initial time of \(t_0 = 0.6\) fm/c, extracted by the JET Collaboration from experimental data on hadron suppression \cite{JET:2013cls}. (a) The gray area represents the variation of $\hat{q}/T^3$ with temperature at the 95\% CL ; (b) the gray area represents the variation of $\hat{q}/T^3$ with temperature at the 68\% CL. The red curve shows the variation of $\hat{q}/T^3$ with temperature at the MAP values.
}
\end{figure*}

In Fig. \ref{qT}, we depict the curve of \(\hat{q}/T^3\) as a function of temperature. The gray area in the figure represents 68\% CL and 95\% CL, while the red curve indicates the MAP values. We observe that the curve reaches a peak above the phase transition temperature and then approaches the value in the pure AdS background. This behavior is distinctly different from the pure AdS background result, where \(\hat{q}/T^3\) remains constant at all temperatures. This indicates that dynamic holographic quantum chromodynamics encodes new features related to the deconfinement phase transition. In Fig. \ref{qTT}, we present the scaled jet transport parameter \(\hat{q} T^3\) as a function of the initial temperature \(T\) for an initial quark jet with energy \(E = 10\) GeV in the most central A+A collisions, with an initial time of \(t_0 = 0.6\) fm/c, extracted by the JET Collaboration from experimental data on hadron suppression \cite{JET:2013cls}. We compare the temperature dependence of \(\hat{q}/T^3\) with this result, and it can be observed that our model's Bayesian inference are in good agreement with the results of the HT-BW model, and they align well with most results at low temperatures.

\begin{figure*}
    \centering
    \includegraphics[width=0.93\textwidth]{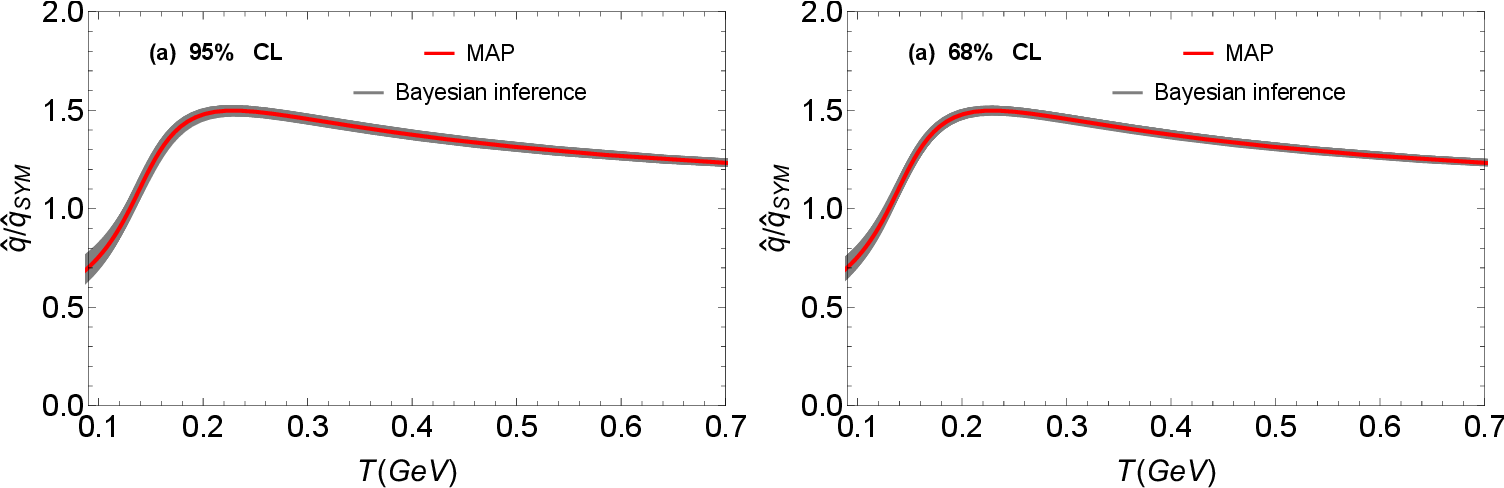}
    \caption{\label{qSYM} The ratio of the jet quenching parameter in the holographic QCD model to \( \hat{q}_{SYM} \) at zero chemical potential. (a) The gray area represents the variation of  \(\hat{q}/ \hat{q}_{SYM} \) with temperature at the 95\% CL; (b) the gray area represents the variation of  \(\hat{q}/ \hat{q}_{SYM} \) with temperature at the 68\% CL. The red curve shows the variation of \(\hat{q}/ \hat{q}_{SYM} \) with temperature at the MAP values. The unit of $T$ is GeV.
}
\end{figure*}

Fig. \ref{qSYM} shows the ratio of the jet quenching parameter in the holographic QCD model to \( \hat{q}_{SYM} \). It can be observed that at lower temperatures, the jet quenching parameter \( \hat{q} \) is below \( \hat{q}_{SYM} \). This implies that, at lower temperatures, the quenching effect of the medium on high-energy jets predicted by the holographic QCD model is weaker than the theoretical predictions in the AdS/Schwarzschild background. However, as the temperature increases, the ratio of \( \hat{q} \) to \( \hat{q}_{SYM} \) first grows, indicating an intensification of the quenching effect, and then starts to decrease after reaching a certain threshold, suggesting a relative weakening of the quenching effect, ultimately approaching 1. This indicates that at higher temperatures, the quenching effect predicted by the holographic QCD model tends to converge with the theoretical predictions made under the AdS/Schwarzschild background.

\begin{figure*}
    \centering
    \includegraphics[width=0.95\textwidth]{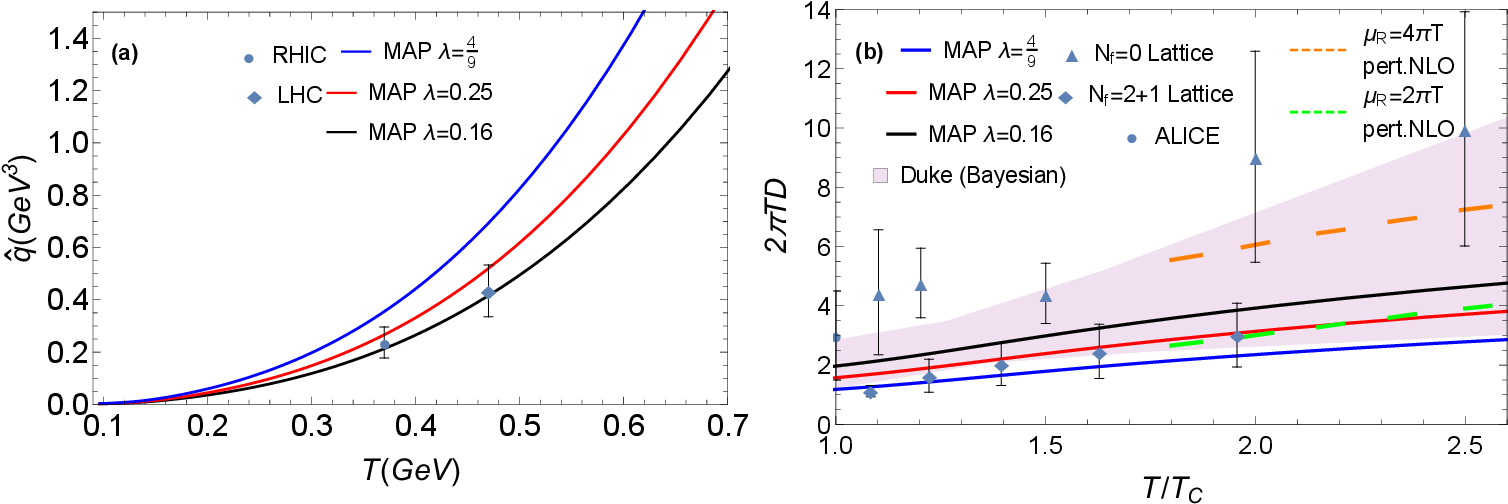}
    \caption{\label{qandD} The figure shows two $\lambda$-sensitive quantities: the jet quenching parameter $\hat q$ (a) and the diffusion coefficient $D$ (b).
}
\end{figure*}

The parameter $\lambda$(or $\alpha'$) is the only new parameter in the paper. At the end of this section, we have added Fig.~\ref{qandD} to illustrate two $\lambda$-sensitive quantities, namely the jet quenching parameter $\hat q$ and the diffusion coefficient. As shown in the figure, increasing the value of $\lambda$ leads to an enhancement of $\hat q$, while the diffusion coefficient decreases as $\lambda$ increases.

\section{Bulk viscosities}
\label{5}
\begin{figure*}
    \centering
    \includegraphics[width=0.92\textwidth]{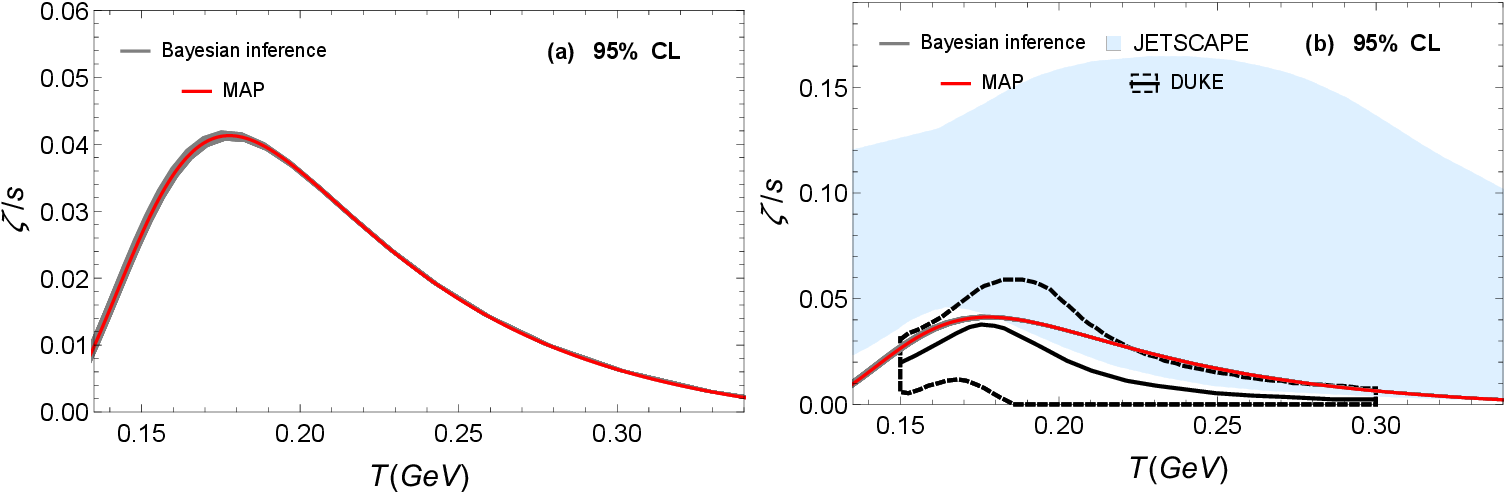}
    \caption{\label{B}   The relationship between \(\zeta / s\) and temperature \(T\) at zero chemical potential. The gray region corresponds to the 95\% CL. The red curves depict the temperature dependence of the bulk viscosity to entropy density ratio based on the MAP values.  The blue region represents the 90\% CL of the JETSCAPE Bayesian model \cite{JETSCAPE:2020shq}, and the black dashed lines and curves indicate the results from the Duke University team \cite{Bernhard:2019bmu}.
}
\end{figure*}

This section will focus on the temperature-dependent behavior of bulk viscosity. We will briefly review how to extract the bulk viscosity from a graviton-dilaton system and present the results graphically.
In the field of four-dimensional field theory, extracting bulk viscosity using the Kubo formula is a viable approach.  
The Kubo formula relates transport coefficients, such as bulk viscosity, to the retarded Green's functions of the stress-energy tensor. Specifically, the bulk viscosity (\(\zeta\)) can be expressed as
\begin{equation}
\zeta = \lim_{\omega \to 0} \frac{1}{\omega} \text{Im}\, G^{R}_{T^{i}_{i}, T^{j}_{j}}(\omega, \mathbf{0}),
\end{equation}
where \(G^{R}_{T^{i}_{i}, T^{j}_{j}}\) is the retarded Green's function of the trace of the spatial stress-energy tensor.  According to the holographic dictionary, the Green's function of the stress tensor can be extracted through metric perturbations (e.g., \(g_{\mu\nu} \to g_{\mu\nu} + h_{\mu\nu}\)). When applying the Kubo formula, we choose the spatial components of the momentum as \(\vec{q} = 0\) and assume that \(h_{\mu\nu}\) depends only on time \(t\) and the \(z\)-direction, i.e., \(h_{\mu\nu} = h_{\mu\nu}(t, z)\). These metric perturbation components should include \(h_{xx}\), \(h_{yy}\), and \(h_{zz}\). We impose spatial rotational symmetry, which allows us to identify the metric perturbations in the three spatial directions as
\(h_{xx}=h_{yy}=h_{zz}\). Following this reasoning, the bulk viscosity can be extracted from the retarded Green's function of the scalar mode. The imaginary part of the retarded Green's function \( G^R(\omega) \) is related to the conserved flux \( \mathcal{F}(\omega) \) via
\begin{equation}
\text{Im}\, G^R(\omega) = -\frac{\mathcal{F}(\omega)}{4\pi G_5}.
\end{equation} 
For the metric of Eq. (\ref{eq2}), the conserved flux takes the form:  
\begin{equation}
\mathcal{F}(\omega) = \frac{L^3 e^{3A(z)} g(z)}{4z^2 (A' - 1/z)^2} \left| \text{Im} \left( h_{xx}^* h_{xx}' \right) \right|.
\end{equation} 
We fix this residual gauge freedom by choosing the dilaton (radial) gauge, i.e. we use the background scalar to parametrize the holographic direction, which is equivalent to setting $\delta\phi=0$.  With this gauge fixing the physical scalar mode is entirely captured by the metric fluctuation $h_{ii}$, and its equation of motion decouples from the remaining perturbations. This procedure follows the standard treatment used, e.g., in Ref.~\cite{Gubser:2008ny} for the computation of bulk viscosity. A notable feature of the perturbation equations of motion is that the equation for the \( h_{xx} \) component decouples completely from \( h_{zz} \) and \( h_{tt} \). The equation of motion for the perturbation \( h_{xx} \) takes the form:  
\begin{equation}
\begin{aligned}
&h_{xx}'' + h_{xx}' [ 3 ( A' - \frac{1}{z} ) + \frac{g'}{g} ] + h_{xx} [ \frac{\omega^2 z^2 e^{-2A}}{L^2 g^2} -
\\& 2 \frac{g'}{g} ( A' - \frac{1}{z} ) - 3 ( A'' - \frac{1}{z^2} ) - 3 ( A' - \frac{1}{z} )^2 ] = 0.
\end{aligned}
\end{equation}
The temperature dependence of the bulk viscosity to entropy density ratio is shown in Figs. \ref{B} (a) and (b), respectively. The pronounced peak of bulk viscosity around $T_c$ could have important consequences for particle spectra and flow observables. In Figs. \ref{B} (a) and (b), 
the gray band displays results derived from posterior parameters within the 95\% CL, while the red curve is obtained from the MAP parameter estimate. The breadth of the gray band directly reflects the uncertainty propagated from the posterior parameter distribution.
It can be observed that the model aligns well with the 90\% CL of the JETSCAPE Bayesian model \cite{JETSCAPE:2020shq} at higher temperatures, as well as the findings from the Duke University team \cite{Bernhard:2019bmu}. 

\begin{figure*}
    \centering
    \includegraphics[width=0.93\textwidth]{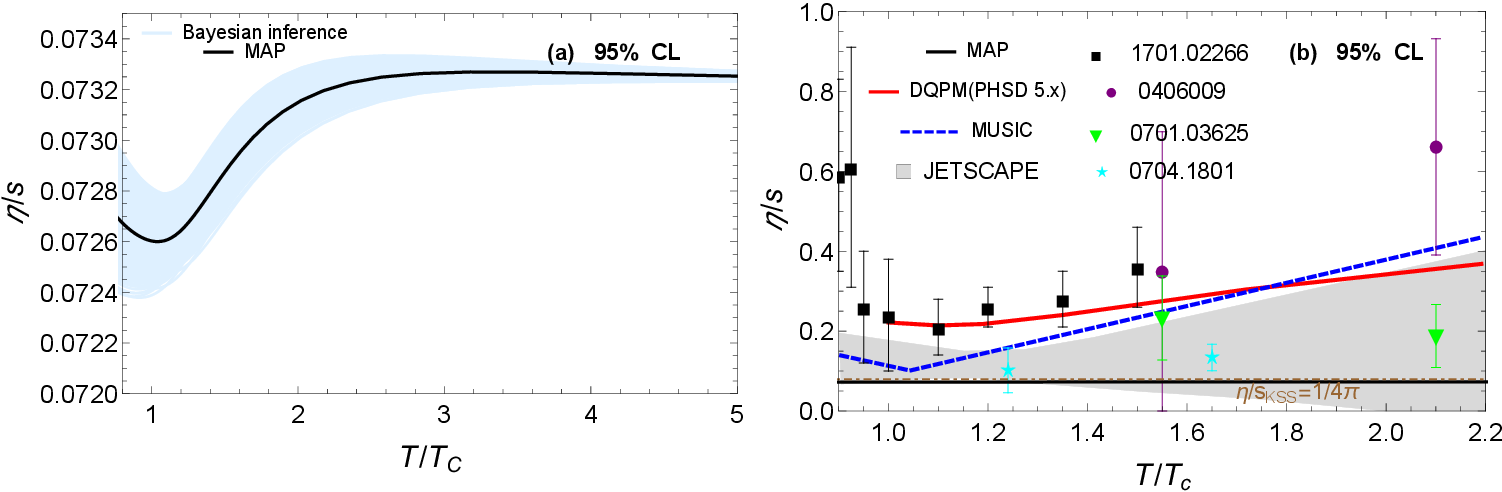}
    \caption{\label{S1}   The functional dependence of the \(\eta/s\) ratio on \(T/T_c\).The blue band represents the 95\% CL, while the black curve shows the functional dependence of the \(\eta/s\) ratio on \(T/T_c\) at the MAP value. The symbols correspond to the LQCD (Lattice QCD) results for pure SU(3) gauge theory: (black squares) \cite{Astrakhantsev:2017nrs}, (green triangles and magenta circles) \cite{Nakamura:2004sy}, and (cyan stars) \cite{Meyer:2007ic}. The gray dash-dotted line represents the Kovtun-Son-Starinets (KSS) bound, \((\eta/s)_{\text{KSS}} = 1/(4\pi)\) \cite{Kovtun:2004de}. The gray band denotes the model-averaged results from the Bayesian analysis of heavy-ion experimental data \cite{JETSCAPE:2020shq}. The red curve corresponds to the results from the DQPM (Dynamical QuasiParticle Model) \cite{Soloveva:2019xph}, while the blue dashed line shows the parameterization of \(\eta/s\) employed in the MUSIC hydrodynamic simulations \cite{Shen:2020jwv}.  
}
\end{figure*}

\section{Shear viscosity with higher derivative corrections}
\label{6}
The shear viscosity coefficient has been extensively studied in the context of Einstein gravity, demonstrating a universal ratio to entropy density \(\eta/s = \frac{1}{4\pi}\) . To introduce temperature dependence in the \(\eta/s\) ratio, higher-derivative gravitational corrections must be incorporated into Einstein gravity \cite{Cremonini:2012ny}. To obtain this temperature dependence, we introduce higher-derivative corrections in the following form:
\begin{equation}\begin{aligned}
&S = \frac{1}{16\pi G_5} \int d^5x \sqrt{-g}
\\&\left[ R - \frac{f(\phi)}{4} F^2 - \frac{1}{2} \partial_{\mu}\phi \partial^{\mu}\phi - V(\phi) + \beta e^{\gamma \phi} R_{\mu\nu\lambda\rho} R^{\mu\nu\lambda\rho} \right].
\end{aligned}
\end{equation}
The $\beta$ is small dimensionless parameter.To compute the shear viscosity coefficient, it is necessary to introduce off-diagonal metric perturbations \( h_{xy} \) to obtain the two-point Green's function of the energy-momentum tensor. The shear viscosity can be calculated via the Kubo formula:  
\begin{equation}
\eta = \frac{1}{9} \lim_{\omega \to 0} \frac{1}{\omega} \text{Im} \, G_{xy,xy}^R (\omega).
\end{equation}
As noted in \cite{Cremonini:2012ny}, \(\eta/s\) depends solely on the background metric at the \(\mathcal{O}(1)\) level. Therefore, we do not consider the \(\mathcal{O}(\beta)\) corrections to the ratio arising from deformations of the background metric. Following the argument in \cite{Cremonini:2012ny}, the calculation of the shear viscosity to entropy density ratio at the \(\mathcal{O}(\beta)\) level can be expressed as:
\begin{equation}
\frac{\eta}{s} = \frac{1}{4\pi} \left[ 1 - \frac{\beta L^2}{c_0} \left( e^{\gamma \phi_h} + \gamma \frac{\phi'(z_h)}{A(z_h)- \frac{1}{z_h}}  e^{\gamma \phi(z_h)} \right) \right],
\end{equation}
Here, \( c_0 = \frac{L^{2} e^{2A_h}}{z_h^{2}} \frac{1}{2 g' A'} = \frac{L^{2} e^{2A_h}}{z_h^{2}} \frac{1}{2 g' (A' - \frac{1}{z_h})} \). When \(\beta = 0.01\) and \(\gamma = -\sqrt{0.3448}\), Fig. \ref{S1} shows the functional dependence of the \(\eta/s\) ratio on \(T/T_c\). It can be observed that the \(\eta/s\) ratio exhibits a dip near \(T = 1.2 T_c\). 

The temperature corresponding to this minimum value almost coincides with the temperature corresponding to the peak of the bulk viscosity. Near the transition region, the trace anomaly exhibits a pronounced peak and the speed of sound $c_s^2$ reaches a minimum, reflecting the strong breaking of conformal symmetry in QCD. These features enhance the interaction strength, which naturally drives $\eta/s$ toward its minimum. This behavior is consistent with general expectations in strongly coupled gauge theories, where $\eta/s$ approaches the KSS bound in the vicinity of the transition.

In Fig. \ref{S1}, 
the blue band represents the results computed using posterior parameters at the 95\% CL, while the black curve illustrates the functional dependence of the \(\eta/s\) ratio on \(T/T_c\) at the MAP value.
The uncertainty associated with the blue band originates from the range of the posterior parameters.
Fig. \ref{S1} (a) displays a magnified view of the results shown in Fig. \ref{S1} (b), where our findings are compared with various model predictions \cite{Astrakhantsev:2017nrs,Nakamura:2004sy,Meyer:2007ic,Kovtun:2004de,JETSCAPE:2020shq,Soloveva:2019xph,Shen:2020jwv}.

\section{Summary}
\label{7}

In this work, we performed a comprehensive investigation of the transport properties of the quark–gluon plasma using a five-dimensional Einstein–Maxwell–dilaton (EMD) holographic model whose parameters are fully constrained by Bayesian inference. In contrast to previous holographic studies that rely on specifically chosen potentials or phenomenological inputs, our model incorporates the complete posterior distributions inferred from lattice QCD thermodynamics, thereby enabling a systematic and statistically controlled quantification of theoretical uncertainties for all transport observables. This Bayesian holographic framework provides a unified and nonperturbative approach to studying the strongly coupled dynamics of QCD matter.

We calculated several key transport coefficients—namely the heavy-quark drag force, the spatial diffusion coefficient, the jet quenching parameter, and the bulk and shear viscosities—at finite temperature and chemical potential. These observables probe complementary aspects of the QGP and jointly capture the interplay between thermal effects, nonconformality, and strong coupling.

The drag force was shown to increase monotonically with both temperature and quark velocity, consistent with a medium whose color charge density and interaction strength grow with temperature. The Bayesian uncertainty band demonstrates that this trend is robust against variations in the EMD model parameters and exhibits quantitative agreement with expectations from strongly coupled dynamics. Because the worldsheet horizon depends sensitively on the background metric, the drag force serves as a stringent probe of the gravitational dual and directly reflects the nonconformal structure encoded in the metric deformation.

For the diffusion coefficient, we demonstrated that the scaled quantity $D/D_{\mathrm{SYM}}$ approaches unity at high temperature, indicating an emergent conformal behavior of the plasma in the $T \gg T_c$ regime. Near the transition region, however, the diffusion coefficient deviates substantially from its conformal value, signaling strong nonperturbative effects. Remarkably, our Bayesian credible intervals align well with lattice QCD calculations, ALICE experimental estimates, and next-to-leading-order perturbative predictions at high temperatures. This agreement demonstrates the predictive power of the Bayesian EMD model.

The jet quenching parameter exhibits a pronounced peak slightly above the phase transition temperature. This behavior mirrors that found in earlier holographic studies but is now obtained with statistically meaningful uncertainty quantification. The peak provides direct evidence of enhanced transverse momentum broadening near $T_c$, where the medium’s color interaction strength is maximal. The magnitude and temperature dependence of $\hat{q}$ derived from our model agree well with phenomenological extractions from RHIC and LHC data, as well as with the HT-BW and HT-M Bayesian extractions from the JET Collaboration.

The viscosities further illuminate the role of strong coupling in the transition region. Consistent with the nonconformal equation of state, we find a sharp peak in the bulk viscosity near $T_c$, accompanied by a minimum in the shear viscosity. The near coincidence of these extrema indicates that the rapid change in the trace anomaly around the phase transition significantly alters the dissipative properties of the medium. Once higher-derivative corrections are included, $\eta/s$ acquires a temperature dependence with a well-defined dip, consistent with expectations from hydrodynamic modeling and with Bayesian extractions of $\eta/s(T)$ from the JETSCAPE and Duke analyses. These results highlight that the EMD model—despite its relative simplicity—captures the essential dynamics underlying both conformal and nonconformal transport effects.

Taken together, our findings demonstrate that a Bayesian-calibrated holographic EMD model can consistently and simultaneously describe a wide range of transport coefficients across the entire temperature regime relevant to heavy-ion collisions. The ability to reproduce lattice QCD thermodynamics, match experimental constraints, and quantify theoretical uncertainties represents a significant advancement over traditional holographic approaches. This work therefore provides a robust and predictive nonperturbative framework for studying QCD matter and offers valuable insights into the microscopic mechanisms governing jet energy loss, heavy-quark dynamics, and viscous transport in the strongly coupled quark–gluon plasma. In the future, our results can be used as inputs to transport models, such as the Linear Boltzmann Transport model, to compute observables like $R_{AA}$, $v_2$, and others \cite{Gubser:2009sn, Walton:1999dy,Bhattacharyya:2024hku}.

\vskip 0.5cm
{\bf Acknowledgement}
\vskip 0.2cm
This work is partly supported by the National Key Research and Development Program of China under Contract No. 2022YFA1604900, by the National Natural Science Foundation of China (NSFC) under Grant No. 12405154, No.12435009 and No. 12275104, and the European Union -- Next Generation EU through the research grant number P2022Z4P4B ``SOPHYA - Sustainable Optimised PHYsics Algorithms: fundamental physics to build an advanced society'' under the program PRIN 2022 PNRR of the Italian Ministero dell'Universit\`a e Ricerca (MUR).

\bibliographystyle{apsrev4-2}
\bibliography{ref}
\end{document}